# The fast X-ray detector system of the FAMU experiment at RAL

**M. Bonesini**, on behalf of the FAMU Collaboration

*Sezione INFN, Dipartimento di Fisica G. Occhialini, Università di Milano Bicocca, Piazza Scienza 3, Milano, 20123, , Italy*

## Abstract

The FAMU experiment at RAL has been designed to study the hyperfine splitting (HFS) of muonic hydrogen and thus measure the Zemach radius of the proton, with a precision better than 1 %. The HFS transition is excited by a tunable MIR laser at ∼ 6790 nm and is recognized by delayed ($\mu$O) X-ray emission around 130-170 keV. The fast X-ray detection system is based on 34 scintillating LaBr$_3$:Ce crystals and one HPGe detector for inter-calibration.

## 1. Introduction

The FAMU (**F**isica degli **A**tomi **Mu**onici) experiment at RAL [1] aims to study the hyperfine splitting (HFS) of muonic hydrogen with the utmost precision. By measuring $\Delta E^{HFS}(\mu^- p)_{1S}$ the Zemach radius $r_Z$ of the proton [2] can be deduced with a precision better than 1 %, The FAMU experiment may shed new light on the problem of the proton radius puzzle [3], where a significant discrepancy was initially found in the measurement of the charge proton radius with impinging muons or electrons.

The experiment utilizes a high-intensity pulsed muon beam, at 55 MeV/c, going to Port 1 of the RIKEN-RAL facility [4]. The muons stop in a hydrogen target, producing muonic hydrogen in a mixture of singlet F=0 and triplet F=1 states. A tunable mid-IR (MIR) pulsed high-power laser [5] is employed to excite the HFS transition of the 1S muonic hydrogen (from F=0 to F=1 states). By exploiting the transfer of muons from muonic hydrogen to another higher-Z gas in the target (such as oxygen), the $(\mu^- p)_{1S}$ HFS transition will be detected by an increase in the number of X-rays from the ($\mu$O) cascade, while tuning the laser frequency $\nu_0$ ($\Delta E_{HFS} = h\nu_0$). The X-rays of interest from oxygen are K$_\alpha$ approximately at 133 keV, K$_\beta$ approximately at 158 keV and K$_\gamma$ approximately at 167 keV. The method used by FAMU [6] is schematically illustrated in Figure 1. Data with the final configuration, including the custom MIR laser, has been collected starting from July 2023: with about 21 days of run in 2023, followed by 17 days in 2024, for a total of 25 different laser frequencies. The laser energy has reached a maximum value of approximately 1.30 ± 0.07 mJ, to be compared with a design request of 1 mJ.

## 2. The FAMU X-ray detection system

The requirements for the muonic X-ray detection system of the FAMU experiment include:

- Good energy resolution at low X-ray energy (∼ 130 − 170 keV) to detect the signal oxygen lines.

- Short signal fall time (less than 200-300 ns) to separate delayed X-rays (signal) from the prompt background.

- Large solid angle coverage at the minimal cost.

Using scintillating crystals, a simple photon readout (such as UBA photomultipliers (PMTs) or SiPM arrays) can be utilized.

The FAMU setup for the 2023-2024 data taking is based on one ORTEC GEM-S5020P4 HPGe for inter-calibrations and 34 LaBr$_3$:Ce detectors arranged in three rings orthogonal to the beamline:

- Six 1" round, 1" thick prototype detectors are read by conventional PMTs [7] [central ring].

- Sixteen 1" round, 0.5" thick detectors are read by Hamamatsu S4161-6050-04-AS SiPM arrays [8] [central ring (6) + upstream ring (10)].

- Twelve 1/2" cubic detectors are read by S13161-3050-04-AS SiPM arrays [9] [downstream ring].

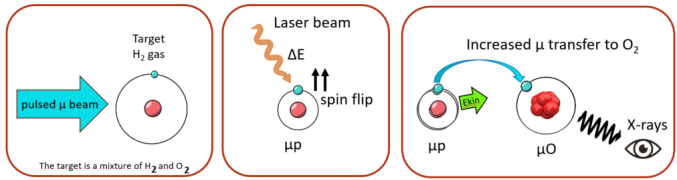

Figure 1: Schematic illustration of the FAMU method to detect hyperfine transitions in muonic hydrogen

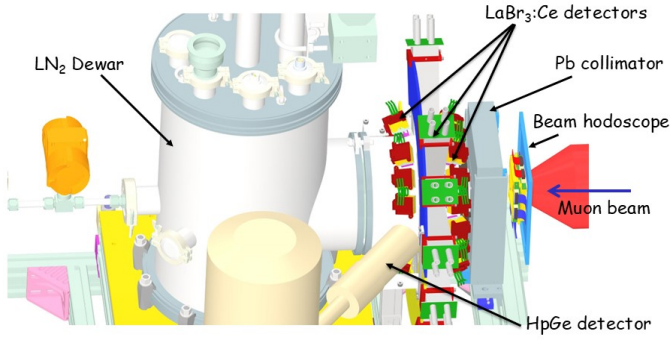

Figure 2: Layout of the FAMU X-ray detection system

In the 2024 data taking the twelve 1/2" detectors of the downstream ring were replaced by 1" round, 0.5" thick detectors, increasing the total solid angle coverage by ∼ 32 %. An enlargement of the layout of the FAMU experimental setup in the region where X-ray detectors are placed is shown in Figure 2. The six 1" detectors with PMT readout are positioned with another six 1" detectors featuring SiPM readout, in the central ring. The detectors with conventional PMT readout have a fully active divider and a custom Digital Pulse Processor (DPP), which uses a 12-bit 500 Ms/s Analog Devices ADC, as explained in [7].

The remaining detectors with an innovative SiPM array readout are described in more detail in the following.

## 3. LaBr$_3$:Ce detectors with SiPM array readout

The mechanics for the LaBr$_3$:Ce detectors with SiPM array readout is realized using a 3D printing, with some elements shown in Figure 3. The flexible design allows for easy reconfiguration of the detectors in use. For example, it was easy to replace the 1/2" cubic detectors used in the 2023 run with 1" round detectors for the 2024 run.

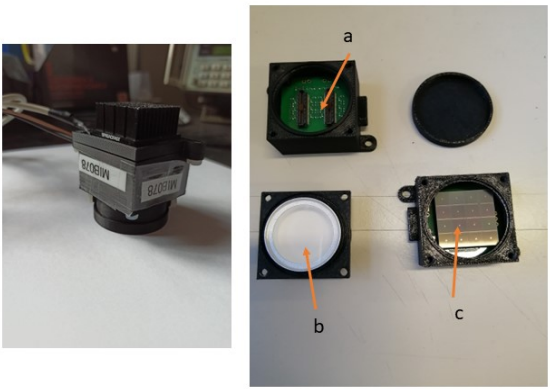

Figure 3: Left panel: image of a complete 1" X-ray detector for the FAMU experiment at RAL, with a passive heat dissipator on the top. Right panel: a) PCB on which the SiPM array is mounted as seen from above, b) holder containing the LaBr3:Ce crystal, c) mounted SiPM array

For the 1/2" LaBr$_3$:Ce detectors with SiPM array readout, conventional parallel ganging was used. For the 1" detectors, an innovative 4-1 circuit was developed with Nuclear Instruments srl, to reduce the signal fall time. In this design, the outputs of four nearby $6 \times 6$ mm$^2$ SiPM are grouped together, followed by an individual pole-zero compensation and amplification. The four sub-array signals are then summed together and inverted to produce a positive signal. This solution results in a temperature increase of 5-7 °C due to the power dissipated by the operational amplifiers Texas Instruments OPA695 used in the circuit (∼ 1 W). To address this problem, a heat dissipator is placed in contact with the PCB's backside via a gap pad in the individual detectors and a fan system for ventilation has been introduced in Port 1.

Good FWHM energy resolution and linearity are obtained, while the pulse fall time is reduced by a factor up to 4X. In the present setup a conservative reduction of a factor 2X is used. Additional details are explained in references [10] and [11] and resumed in table 1.

The output signal, resulting from the summed analog signals of the 16 cells of a SiPM array, is then fed into a CAEN V1730 FADC [1] without further amplification. Laboratory measurements, done at INFN Milano Bicocca, for all the detectors with SiPM array readout are shown in Figure 4. The detectors' response is linear in the range 100-1200 keV and the FWHM resolution around 3 % ( 8 %) at 662 (120) keV. All results have been obtained by testing the detectors inside a climatic chamber at 20 °C, to simulate the temperature at Port 1 of RIKEN RAL.

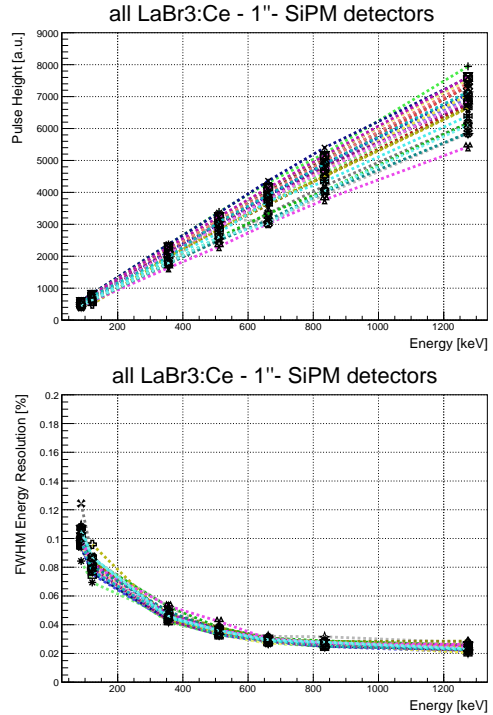

Figure 4: Top panel: linearity for the full sample of 1" LaBr$_3$:Ce detectors with SiPM array readout. used in the FAMU experiment Bottom panel: FWHM energy resolution for the same sample of detectors.

---

[1] 500 MHz bandwidth, 14 bit resolution
2

The SiPM gain drift with temperature can be corrected on-line using a NIM custom module (8 channels) based on CAEN A7585D digital power supplies, with temperature feedback. Temperature is measured using a TMP37 sensor from Analog Devices, placed in thermal contact with the backside of the PCB on which the SiPM arrays are mounted. The variation in the $Cs^{137}$ photopeak pulse height was reduced from 41 % to 5 % for S14161-6050-AS arrays, scanning the temperature range $10 - 30\,^0C$. For further details see references [12], [13].

## 4. X-ray detectors performances

The LaBr$_3$:Ce 1" detectors with SiPM readout exhibited an increased dark current, with respect to laboratory measurements, during data taking, as reported in Figure 5. This increase is likely due to the SiPM damage induced by a low-rate production of neutrons by $\mu$-Au interactions in the silver moderator within the target. This issue was solved by a 24 hours baking at 125 $^0$C of the involved SiPM arrays, as shown in Figure 6.

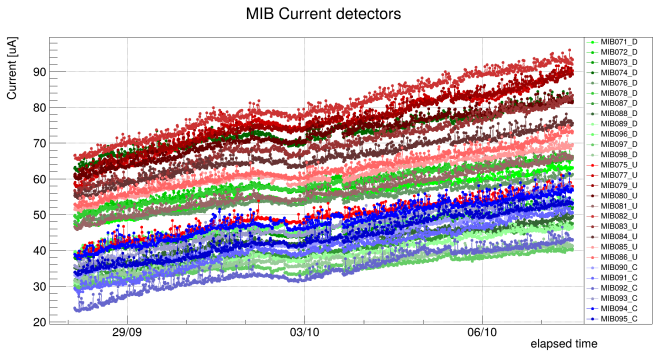

Figure 5: Dark currents for all the 1" LaBr3:Ce detectors read by SiPM arrays, during the September/October 2024 run (no beam on 2-3 October).

All LaBr$_3$:Ce detectors were calibrated, prior to each data-taking period, with $Am^{241}$, $Ba^{133}$ and $Cs^{137}$ sources. For every batch of data, each detector was further calibrated in-situ using known peaks from C, Pb and Ag, with an automatic procedure, shown in Figure 7. A parabolic fit for the ADC vs Energy relationship was used, to accomodate detectors with either the PMT readout (less linear) or the SiPM readout.

The timing and FWHM energy resolution of the 4 types of detectors are shown in Table 1. FWHM energy resolutions at $^{137}$Cs (662 keV) and $^{57}$Co (120 keV) peaks are measured in laboratory, while results at 142 keV (Ag peak) are from beam data at RAL with 55 MeV/c impinging muons. For comparison, the FWHM energy resolution at the 142 keV muonic silver peak is 1.26 ± 0.17% from the HPGe detector, at the cost of a much longer fall time.

Figure 8 shows the signal oxygen lines after H$_2$ background subtraction, for a H$_2$+O$_2$ (1.5 % wt.) gas mixture inside the target, as detected by the LaBr$_3$:Ce detectors [2]. Data were collected in December 2023, with 55 MeV/c impinging muons.

---
[2]negative values at the beginning are due to an imperfect background subtraction

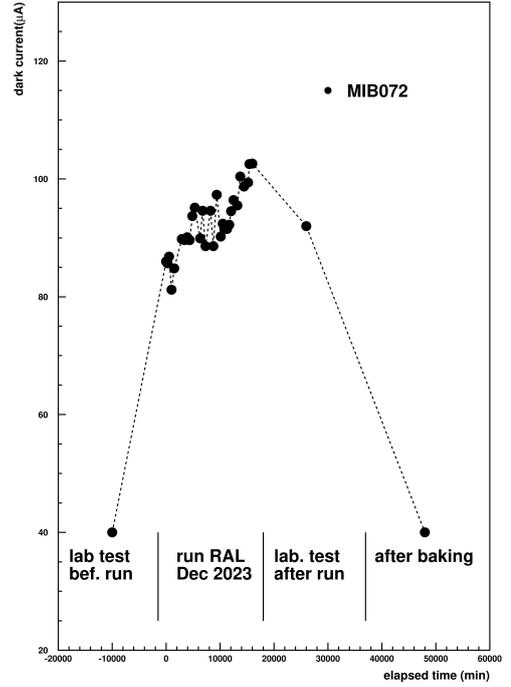

Figure 6: Behaviour of dark currents vs time for the worst LaBr3:Ce detector with SiPM array readout. Before the December 2023 run: laboratory tests, during the December 2023 run, after the run: laboratory tests and after baking at 125 C for 24 hours.

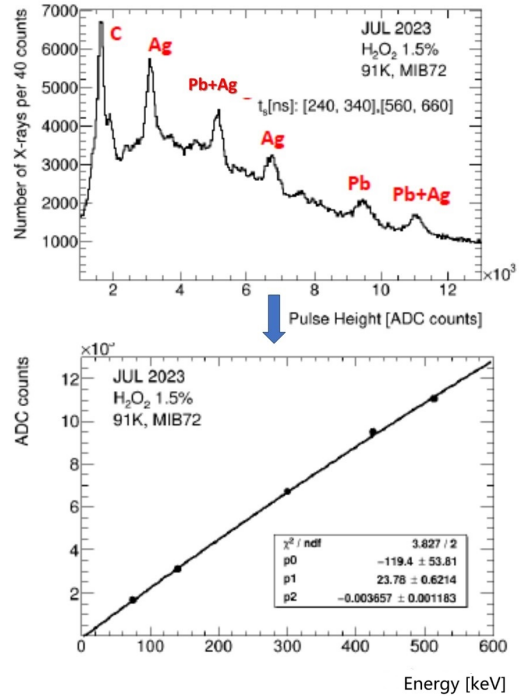

Figure 7: In-situ calibration procedure from ADC counts to energy in keV for a LaBr$_3$:Ce detector. Top panel: considered peaks in data taking for calibration. Bottom panel: result of the calibration procedure (the first Pb+Ag peak is not used).

Table 1: Average performances of FAMU X-ray detectors. Rise time and fall time (10-90 %) in ns refer to detectors' summed analog outputs. FWHM energy resolutions (R) are in percent. Sample means and RMS deviations are reported.

|           | 1" - PMT   | 1" - SiPM    | 1/2" - SiPM   | HpGe        |
|-----------|------------|--------------|---------------|-------------|
| rise time | 14 ± 1     | 29.3 ± 1.5   | 42.8 ± 4.7    | 150         |
| fall time | ∼ 60       | 147.1 ± 12.8 | 372.4 ± 17.4  | 1000-2000   |
| R 662 keV | 3.5-4.6    | 2.94 ± 0.14  | 3.27 ± 0.11   | .442 ± .001 |
| R 120 keV | 7.2-8.1    | 8.03 ± 0.39  | 8.44 ± 0.63   | 1.010 ± .001|
| R 142 keV | 12.3 ± 1.2 | 7.0 ± 0.3    | 7.5 ± 0.3     | 1.26 ± 0.17 |

No distinction is made for the laser/no-laser data. Clear $K_\alpha$ and $K_{\beta/\gamma}$ signal lines are seen.

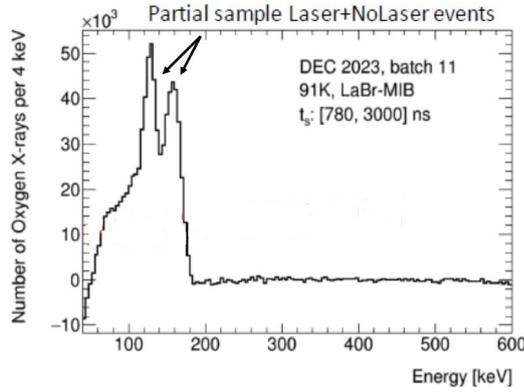

Figure 8: $K_\alpha$, $K_{\beta,\gamma}$ signal oxygen lines, as seen by LaBr$_3$:Ce detectors.

## 5. Conclusions

To ensure clear detection of the $K_\alpha$, $K_{\beta/\gamma}$ signal lines from $\mu$O, a fast X-rays detection system based on LaBr$_3$:Ce scintillating crystals is used. The SiPM array readout performs comparably to the one with PMT readout. A significant reduction in the signal fall time has been obtained with a dedicated 4-1 design of the SiPM array readout.

**Acknowledgements**

The skilful help of the INFN MIB mechanics workshop, in particular of R. Gaigher, is acknowledged. In addition, I would like to thank Francesco Caponio and Andrea Abba from Nuclear Instruments srl for many interesting discussions on SiPM readout and M. Saviozzi from CAEN srl for help on electronics.